\documentclass[
	aps,
	pra,
	twocolumn,s
	uperscriptaddress,
	nofootinbib,
	amsmath,amssymb
]{revtex4-2}

\usepackage[T1]{fontenc}
\usepackage{bm}
\usepackage{upgreek}
\usepackage[dvipsnames,table,xcdraw]{xcolor}
\usepackage{graphicx}
\graphicspath{{fig/}}
\usepackage{booktabs}
\usepackage{array}
\usepackage[normalem]{ulem}
\usepackage{hyperref}
\hypersetup{
    colorlinks=true,
    citecolor=blue,
    linkcolor=blue,
    filecolor=magenta,
    urlcolor=blue
}

\newcommand{\ii}{\mathrm{i}}
\newcommand{\dd}{\mathrm{d}}

\begin{document}

\title{Enhanced chiral response of hybrid photonic structures empowered by Mie resonances}

\author{Xingguang Liu}
\email{liu\_xg@hit.edu.cn}
\affiliation{School of Physics, Harbin Institute of Technology, Harbin 150001, China}
\affiliation{Heilongjiang Provincial Key Laboratory of Advanced Quantum Functional Materials and Sensor Devices, Harbin 150001, China}

\author{Wenzhe Hao}
\affiliation{School of Physics, Harbin Institute of Technology, Harbin 150001, China}

\author{Ivan Toftul}
\email{ivan.toftul@anu.edu.au}
\affiliation{Research School of Physics, Australian National University, Canberra ACT 2601, Australia}

\author{Junqing Li}
\affiliation{School of Physics, Harbin Institute of Technology, Harbin 150001, China}

\author{Yongkang Dong}
\affiliation{National Key Laboratory of Laser Spatial Information, Harbin Institute of Technology, Harbin 150001, China}

\author{Yuri Kivshar}
\email{yuri.kivshar@anu.edu.au}
\affiliation{Research School of Physics, Australian National University, Canberra ACT 2601, Australia}

\date{\today}

\begin{abstract}
Chiroptical response provides a powerful route to control light-matter interactions through light handedness, with crucial importance in polarization control, chiral sensing, and nonlinear photonics. Hybrid structures may offer a novel route to engineer such responses by coupling plasmonic field localization with dielectric Mie resonances within a single nanostructure. However, in chiral metasurfaces, hybridization has largely been treated as a means of resonance enhancement, while its role in actively reshaping and reversing intrinsic plasmonic chirality remains largely unexplored. Here, we demonstrate a hybrid metasurface in which a dielectric Mie resonance mediates enhanced chiral response of plasmonic building blocks. By coupling a Si nanocylinder to trapezoidal Au nanobars, a weak and mode-selective plasmonic chiral seed is reorganised into two opposite-handed hybrid resonances, producing near-unity circular dichroism with spectral sign flipping. Extending this mechanism to nonlinear regime, the same hybrid-mode selectivity produces strongly nonlinear chiral emission and near-unity third-harmonic generation circular dichroism of either sign. We believe these results establish plasmonic-Mie hybridization as a novel mechanism for engineering high-purity, sign-switchable linear and nonlinear chirality in compact metadevices.
\end{abstract}


\maketitle

Chirality is one of the key concepts in physics and chemistry. 
Optical chirality is defined through a dissimilar interaction with left and right circularly polarized plane waves, often referred to as \textit{circular dichroism} (CD)~\cite{Tang2010OpticalChirality,Lininger2023Chirality,Deng2024ChiralMetasurfaces,Mun2020ElectromagneticChirality}.
Controlling optical chirality at subwavelength scales is essential for polarization optics, chiral sensing, spin-selective light-matter interaction, and nonlinear photonics~\cite{Lininger2023Chirality,Deng2024ChiralMetasurfaces,Hu2020UVCircularDichroism,Mohammadi2021DualNanoresonators,Koshelev2023ResonantChiral,Jiang2026SuperchiralSensing,Both2022ChiralSensingMechanism,Warning2021ChiralitySensing,Bai2021PolarizationConversion}. 
Metasurfaces provide an attractive route to enhance the chiral response by engineering resonant electromagnetic modes, local field distributions, and symmetry properties within deeply subwavelength unit cells.
A major goal in chiral nanophotonics is to approach the so-called maximum chiral response, where one circular polarized wave is almost completely suppressed while the opposite handedness wave is transmitted or scattered efficiently~\cite{FernandezCorbaton2016MaximumChirality,Gorkunov2020BICMaximumChirality,Gorkunov2021NearLosslessChirality,Shi2022PlanarChiralBIC,Kuehner2023OutOfPlaneChirality,ACS2024TopologicalChiralBIC,Kumar2025MaximalBilayer}. Several recent studies have demonstrated that maximum chirality can be realized by carefully engineering resonant metasurfaces. One approach employs tilted or symmetry-broken resonators operating in nonlocal regimes associated with bound states in the continuum (BICs), where high-Q resonances strongly enhance the interaction between circularly polarized light and chiral modes~\cite{Gorkunov2020BICMaximumChirality,Shi2022PlanarChiralBIC,Chen2023IntrinsicChiralBIC,Kuehner2023OutOfPlaneChirality,Kim2024NonlocalStokes,Zhu2026DualBandChirality,Hu2026RobustChirality,Overvig2021ChiralQuasiBIC,Zhang2022ChiralEmission,Shakirova2025CrosstalkingQuasiBICs}. Another strategy uses multipolar effects in binary or bilayer photonic structures, for example, rotated C$_4$-symmetric apertures, where strong mode coupling produces resonant chiral photonic modes with nearly perfect circular selectivity~\cite{Tanaka2020ChiralBilayer,Kumar2025MaximalBilayer,Kumar2026IntrinsicChiralModes,Gromyko2024TwistedBilayer}. However, most existing strategies rely on deliberately engineered chiral photonic modes, nonlocal resonances, or multilayer geometric coupling~\cite{Deng2024ChiralMetasurfaces,Shi2022PlanarChiralBIC,Chen2023IntrinsicChiralBIC,Kuehner2023OutOfPlaneChirality,Kim2024NonlocalStokes,Kumar2025MaximalBilayer,Kumar2026IntrinsicChiralModes,Tang2026BilayerNonlocal,Khaliq2023ChiralMetasurfacesReview,Gryb2023SimpleRotations,Mohammadi2023ChiralityTransfer}. A different and less explored question is whether maximum chirality can emerge from the hybridization between two physically distinct modal systems, rather than from structure alone.

\begin{figure*}[t]
	\centering
	\includegraphics[width=0.8\linewidth]{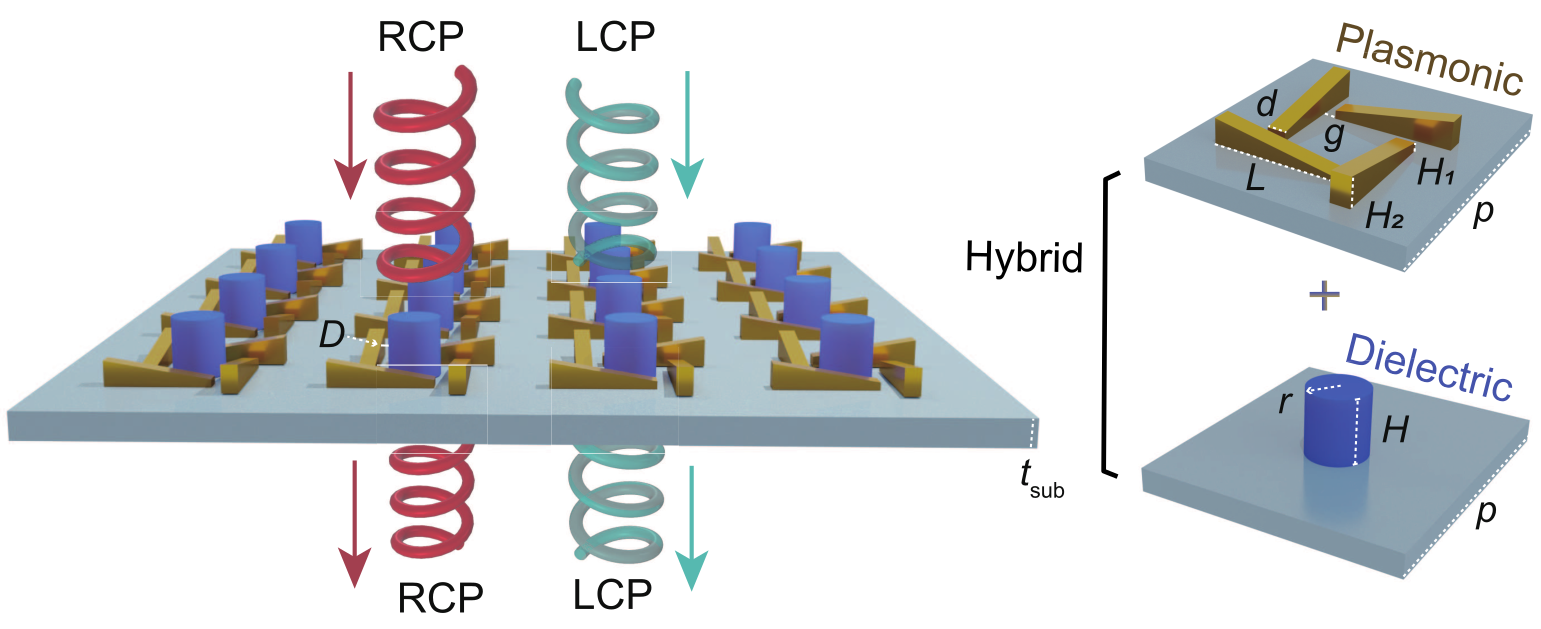}
	\caption{\textbf{Hybrid metasurface design.} Left, the hybrid metasurface under circularly polarized illumination. Four trapezoidal Au nanobars surround a Si nanocylinder on a free-standing silica membrane of thickness $t = 200$~nm, with period $p = 1150$~nm (square lattice), and bar--cylinder edge-to-edge distance $D = 54$~nm. The unit cell is C$_4$ symmetric about the vertical axis, which forbids circular polarization conversion at normal incidence and preserves the transmitted handedness, yet it has no mirror plane thus chiral. Right, the two building blocks: plasmonic and dielectric designs. The Au bars have $L=600$~nm, $d = 35$~nm, gap $g = 125.4$~nm, and heights $H_1 = 50$~nm and $H_2 = 200$~nm; the Si cylinder has radius $r = 150$~nm and height $H = 255$~nm.}
	\label{fig:concept}
\end{figure*}

Here we uncover a novel mechanism of near-unity circular dichroism that can be realized in hybrid plasmonic-dielectric structures. 
Hybridization of resonances is a novel avenue to take advantage of both worlds and overcome the inherent limitations of individual systems, enabling advanced functionalities and applications~\cite{Zhang2026EngineeringHybrid,Guan2022HybridMaterial,Barreda2022HybridNanostructures,Dmitriev2023HybridNanoantenna,Randerson2024HybridMiePlasmonic,Tang2026BilayerNonlocal,Fu2023TaleTwoResonances,Oleynik2024HybridLatticeCoupling}. 
As an example, here we consider hybrid plasmonic-Mie resonator structure that synergistically combines strong near-field enhancement of plasmonic components with the low-loss, multipolar resonances of dielectric Mie nanoparticles~\cite{Guo2016MultipolarCoupling,Barreda2022HybridNanostructures,Babicheva2024MieMetaphotonics,Martens2022OnsetChirality,Ray2020HybridSensing,McPolin2023HybridGoldSilicon,Dmitriev2023HybridNanoantenna,Randerson2024HybridMiePlasmonic,Mohammadi2021DualNanoresonators,Serrera2024AmplifiedChiralSensing}. 
We reveal that, compared to the ‘plasmonic-only’ structure, the hybrid structure with the Si nanocylinder leads to a pronounced reshaping of the CD spectrum. In particular, the initially moderate single CD peak is transformed into two near-unity chiroptical responses of opposite handedness. 
This CD splitting naturally gives rise to a spectral chiral flipping, with opposite handedness emerging at the two resonance wavelengths.
We further extend this hybridization mechanism into the nonlinear regime. The resulting third-harmonic generation circular dichroism (THG-CD) reaches near-unity values over finite spectral windows before switching sign between the two hybrid modes.
These behaviours cannot be understood as a simple superposition of resonances. Instead, the Mie resonance acts as an active modal mediator that perturbs, enhances, and reorganises the intrinsic plasmonic chirality. We believe our work provides a new route to engineer how plasmonic localization and dielectric multipolar scattering interfere, and therefore offers a natural platform for enhancing chiroptical responses, opening opportunities for chiral sensing, nonlinear frequency conversion, and polarization-selective metadevices.

Figure~\ref{fig:concept} frames the central concept of this work by introducing the  plasmonic--dielectric hybrid metasurface. 
The right panel conceptually decomposes the two elementary components:  a plasmonic (made of Au) assembly and a dielectric (made of Si) nanocylinder. The metal component, composed of four trapezoidal nanobars with length $L$, height parameters $H_1$, $H_2$ and side dimension $d$ supports localized plasmonic modes and provides an intrinsic chiral scattering background due to its broken symmetry. The dielectric component, defined by the cylinder radius $r$ and height $H$, possesses a typical Mie resonant character. The gap between two gold bars is $g$. The left schematic shows the full hybrid metasurface by integrating these two building blocks within the same unit cell on a free-standing silica membrane. 
The period of the metasurface unit cell is denoted as $p$. And the edge-to-edge normal distance between the gold bars and silicon cylinder is $D$. This configuration brings the localized plasmonic modes and the dielectric Mie modes into close spectral and spatial proximity, allowing their coupling to reshape the response of the full meta-atom. Under circularly polarized illumination, the hybrid structure therefore does not simply behave as the sum of a plasmonic resonator and a dielectric resonator. Instead, it provides a complex mode coupling in which plasmonic localization, dielectric multipolar scattering, vertical asymmetry, and circular-polarization-dependent interference can act together.

\begin{figure}[htb]
	\centering
	\includegraphics[width=\linewidth]{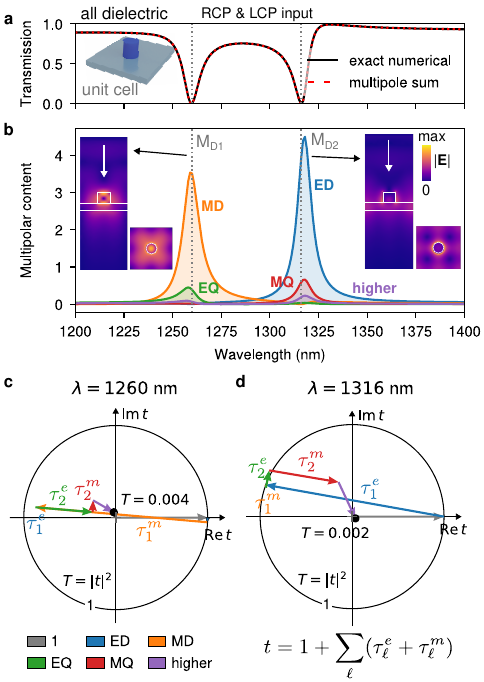}
	\caption{\textbf{All-dielectric achiral design.}
	(a) Transmission of the dielectric-only achiral metasurface. Solid line, full-wave result; dashed line, multipole sum \eqref{eq:t_multipole}. The dips locate the resonances M$_{\rm D1}$ (1260~nm) and M$_{\rm D2}$ (1316~nm). 
	(b) Multipolar content \eqref{eq:multipole_content}, split into dipole (ED, MD), quadrupole (EQ, MQ), and summed $\ell \geqslant 3$ (higher) terms. M$_{\rm D1}$ is dominated by  a magnetic dipole, while M$_{\rm D2}$ is dominated by an electric dipole. Insets, $|\mathbf E|$ in vertical and horizontal cuts at the two resonances, with the arrow marking the incidence direction.
	(c,d) Illustration of the sum in transmission coefficient $t$ in the complex plane, Eq.~\eqref{eq:t_multipole}.  At both resonances the multipolar radiation nearly cancels the direct term, leaving $T = |t|^2 \approx 0$.}
	\label{fig:dielectric}
\end{figure}

To uncover the origin of the Mie-assisted enhancement mechanism, we first isolate the dielectric contribution by examining the ‘dielectric-only’ structure, as shown in Figure~\ref{fig:dielectric}.  %
As expected from the preserved symmetry, the transmission spectra of the dielectric-only structure under left- and right- circularly polarized (LCP and RCP) are identical (Figure~\ref{fig:dielectric}a). 
In this work we consider only four-fold symmetric structures for which cross conversion transmission is forbidden, $T_{\text{RL}} = T_{\text{LR}} = 0$. 
To quantify linear CD we calculate its co-polarized counterpart, which has a clear link to the geometrical definition of the chirality~\cite{Toftul2024,Kumar2025MaximalBilayer}
\begin{equation}
    \mathrm{CD}=
    \frac{
    T_{\mathrm{LL}}-T_{\mathrm{RR}}
    }{  T_{\mathrm{LL}}+T_{\mathrm{RR}}  },
    \label{eq:linear_CD}
\end{equation} where the first/second subscript denote the analysed output/incident pump polarization state, respectively. Therefore, $T_{\mathrm{LR}}$ represents right-hand circular polarization (RCP) incidence with left-hand circular polarization (LCP) output, and $T_{\mathrm{RL}}$ represents LCP incidence with RCP output. Figure~\ref{fig:dielectric}a confirms that the isolated dielectric part does not intrinsically support chiral optical response in the present geometry.
Nevertheless, high-index Si nanocylinders represent a canonical platform for spectrally distinct Mie resonances. %

To characterize the multipolar modal composition of the periodic metasurface, we perform a vector-spherical-harmonic (VSH) decomposition of the polarization current induced within one unit cell. The current is extracted from the self-consistent periodic full-wave solution as
\begin{equation}
\mathbf J(\mathbf r)=
-{\rm i}\omega\varepsilon_0
\left[
\varepsilon_r(\mathbf r,\omega)-\varepsilon_h
\right]
\mathbf E(\mathbf r),
\label{eq:Jlinear}
\end{equation}
Here $\mathbf E$ is the total field which is obtained numerically with Floquet periodic boundaries for RCP or LCP incident helicity. The resulting multipole coefficients describe the lattice-dressed response of the meta-atom rather than that of an isolated scatterer, thus no lattice sum needed compared to Ref.~\cite{Rahimzadegan2022}.

We denote the VSH as $\mathbf N_{m \ell}(k\mathbf r)$ (electric) and $\mathbf M_{m \ell}(k\mathbf r)$ (magnetic). The radiation associated with the unit-cell current \eqref{eq:Jlinear} is expanded in normalized electric- and magnetic-type VSH as~\cite{Jackson1998,Alaee2018,Grahn2012,Rahimzadegan2022}
\begin{equation}
	\mathbf E_{\rm sc}(\mathbf r)=
	E_0
	\sum_{\ell=1}^{\infty}
	\sum_{m=-\ell}^{\ell}
	\left[
	a_{m \ell}\mathbf N_{m \ell}^{(3)}(k\mathbf r)
	+
	b_{m \ell}\mathbf M_{m \ell}^{(3)}(k\mathbf r)
	\right],
	\label{eq:expansion}
\end{equation}
where $E_0$ is the dimensional constant equal to the amplitude of the incident plane wave,
$a_{m \ell}$ and $b_{m \ell}$ are the electric- and magnetic-type spherical multipole coefficients, respectively, and superscript ``$(3)$'' denotes the outgoing spherical-Hankel radial dependence. Their amplitudes are obtained by projecting the polarization current \eqref{eq:Jlinear} onto the corresponding regular VSHs~\cite{Jackson1998,Toftul2026thesis}: $a_{m \ell} = - \frac{k^2}{\ell (\ell + 1)} \frac{\sqrt{\mu_0}}{E_0 \sqrt{\varepsilon_{0} \varepsilon_h }}\int \mathbf{J} \cdot \mathbf{N}_{m\ell}^{(1)*} \dd V$ and $b_{m \ell} = - \frac{k^2}{\ell (\ell + 1)} \frac{\sqrt{\mu_0}}{E_0 \sqrt{\varepsilon_{0} \varepsilon_h }}\int \mathbf{J} \cdot \mathbf{M}_{m\ell}^{(1)*} \dd V$. 
In the sub-diffractive regime at normal incidence, the outgoing wave is a plane wave and is described by a single complex transmission coefficient, so that the transmitted field is $t \mathbf{E}_{\text{inc}}$. We express $t$ through the multipole coefficients of Eq.~\eqref{eq:expansion} as~\cite{Toftul2026thesis}
\begin{align}
\begin{split}
    t & = 1 + \sum_{\ell = 1}^{\infty} \left( \tau^{e}_{\ell} + \tau^{m}_{\ell}\right), \\
    \tau^{e}_{\ell} & = C_{\ell} \sum_{m = \pm 1} m \, a_{m\ell}, \quad 
    \tau^{m}_{\ell} = - C_{\ell} \sum_{m = \pm 1} m^2 b_{m\ell}.
\end{split}
\label{eq:t_multipole}
\end{align}
Here $C_{\ell} = 2\pi (kp)^{-2} \ii^{\ell+1}\sqrt{\ell(\ell + 1)(2\ell + 1) / (8\pi)}$ and $p$ is the lattice period.
Only the $m = \pm 1$ multipoles radiate along the surface normal. 
The coefficients with $|m| \neq 1$ are in general nonzero, but they do not couple to the zeroth diffraction order and therefore drop out of Eq.~\eqref{eq:t_multipole}.
We resolve the individual contributions through the quantities
\begin{equation}
\begin{gathered}
    \text{ED} = |\tau^{e}_{1}|^2, \;\;
    \text{MD} = |\tau^{m}_{1}|^2, \;\;
    \text{EQ} = |\tau^{e}_{2}|^2, \;\;
    \text{MQ} = |\tau^{m}_{2}|^2, \\
    \text{higher} = \left|\sum_{\ell = 3}^{\ell_{\text{max}}} \left( \tau^{e}_{\ell} + \tau^{m}_{\ell}\right) \right|^2, 
\end{gathered}
\label{eq:multipole_content}
\end{equation}
retaining their relative phases in the complex-plane construction of Eq.~\eqref{eq:t_multipole}.
Details of the VSH multipole decomposition and numerical implementation are provided in Supporting Information Note2 and Note4.

\begin{figure*}[t]
	\centering
	\includegraphics[width=\linewidth]{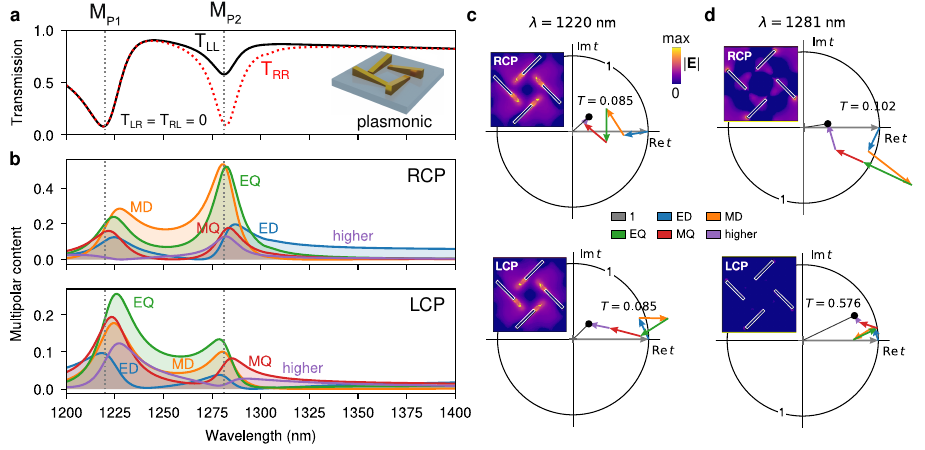}
	\caption{\textbf{Plasmonic chiral design.}
	(a) Co-polarized transmission of the plasmonic-only chiral metasurface, $T_{\rm LL}$ (solid) and $T_{\rm RR}$ (dotted), with $T_{\rm LR} = T_{\rm RL} = 0$ by the C$_4$ symmetry. The two channels coincide at M$_{\rm P1}$ (1220~nm) and separate at M$_{\rm P2}$ (1281~nm), where ${\rm CD} = +0.70$.
	(b) Multipolar content \eqref{eq:multipole_content} under RCP (top) and LCP (bottom) incidence. Both resonances have no dominant multipoles. The input helicity changes their relative weights and phases. 
	(c,d) Complex-plane construction as in Figure~\ref{fig:dielectric}c,d, at M$_{\rm P1}$ (c) and M$_{\rm P2}$ (d) for RCP (top) and LCP (bottom). The two helicities reach the same $T$ along different multipolar paths at M$_{\rm P1}$, but not at M$_{\rm P2}$. Insets, top view of the electric near field, with handedness-dependent hot spots at the bar ends at M$_{\rm P2}$.}
	\label{fig:plasmonic}
\end{figure*}

As shown in Figure~\ref{fig:dielectric}b, the VSH multipolar decomposition resolves two spectrally distinct resonances of the Si nanocylinder on the silica membrane. The shorter-wavelength mode, M$_{\rm D1}$, is dominated by the magnetic-dipole (MD) component, with weaker contributions from ED and higher-order multipoles, whereas the longer-wavelength mode, M$_{\rm D2}$, is predominantly electric-dipolar (ED), accompanied by minor magnetic and higher-order components. The $|\mathbf E|$ maps in the insets of Figure~\ref{fig:dielectric}b, computed at normal incidence under a circularly polarized plane wave, support these assignments. At M$_{\rm D1}$ the electric field wraps around the cylinder and leaves a node at its centre, as expected for the circulating displacement current of a magnetic dipole, whereas at M$_{\rm D2}$ it is expelled from the cylinder interior and peaks at its rim.
The multipolar spectra under LCP and RCP illumination coincide, indicating that the two incident helicities excite identical multipolar compositions and amplitudes at both resonances. This helicity-independent internal response is fully consistent with the overlapping LCP/RCP transmission spectra in Figure~\ref{fig:dielectric}a and the resulting zero CD (Eq.~\eqref{eq:linear_CD}). Figures~\ref{fig:dielectric}c and~\ref{fig:dielectric}d trace the two transmission dips back to this multipolar content. At both resonances the multipolar radiation almost exactly cancels the direct term of Eq.~\eqref{eq:t_multipole}, so that $T \simeq 0$. We note that while MD and ED look dominant in Fig.~\ref{fig:dielectric}b, the role of other multipoles in sum~\eqref{eq:t_multipole} is not negligible.

We then turn to the ``plasmonic-only'' structure to isolate the intrinsic chiral modes of the Au nanobars. 
Figure~\ref{fig:plasmonic} shows the intrinsic chiroptical response of the plasmonic building block. As seen in Figure~\ref{fig:plasmonic}a, the two co-polarized transmission channels separate around only one of the two resonances, M$_{\rm P2}$ (1281~nm), where Eq.~\eqref{eq:linear_CD} gives ${\rm CD} = +0.70$. This indicates that the plasmonic component carries a finite chiral response due to its broken symmetry, but this response is concentrated in a specific localized plasmonic mode rather than distributed across all resonances.
The same panel identifies the transmission pathways responsible for this CD. Around M$_{\rm P2}$, $T_{\rm LL}$ and $T_{\rm RR}$ exhibit a clear difference in spectral weight, producing a pronounced positive CD, whereas the cross-circular components $T_{\rm LR}$ and $T_{\rm RL}$ vanish over the entire spectral range. This behaviour is dictated by the in-plane C$_4$ rotational symmetry under normal incidence, which preserves handedness.
Near M$_{\rm P1}$ (1220~nm), the two co-circular channels overlap, resulting in the almost vanishing CD.

The near-field distributions shown in the insets of Figures~\ref{fig:plasmonic}c and~\ref{fig:plasmonic}d provide a real-space picture corresponding to the CD spectrum. At M$_{\rm P1}$, LCP and RCP excitation produce weak and broadly similar electric-field patterns, with localized fields mainly confined near the inner edges of the Au nanobars. By contrast, at M$_{\rm P2}$ a pronounced handedness-dependent field localization appears, with RCP excitation generating much stronger hotspots at the bar ends than LCP excitation. This spatial asymmetry mirrors the transmission imbalance.

The multipolar analysis in Figure~\ref{fig:plasmonic}b provides a modal understanding of the two plasmonic resonances. 
Neither M$_{\rm P1}$ nor M$_{\rm P2}$ can be assigned to a single elementary multipole. Both resonances instead spread over a comparable set of electric- and magnetic-type spherical harmonics, in which dipole and quadrupole channels carry similar weight and none dominates.  Capacitive coupling across the gaps and the nonuniform current flow along the vertically asymmetric bars form effective current loops that feed the magnetic channels, while the broken vertical symmetry admixes the electric and higher-order ones.
The two resonances, however, exhibit markedly different helicity dependence. 
Around M$_{\rm P1}$, both LCP and RCP excitations access the same set of multipolar channels, although the corresponding amplitudes are not identical. Importantly, these coefficients characterize the internal multipolar content of the driven unit-cell response and therefore cannot be directly identified with the helicity-dependent far-field transmission. The observable CD is determined by the coherent radiation of all excited multipolar components into the open transmission channels, including their relative amplitudes, phases, and radiation patterns. Figure~\ref{fig:plasmonic}c makes this explicit. The two helicities follow visibly different paths in the complex plane, yet their vector sums terminate at the same $T$, so that $T_{\rm LL} \simeq T_{\rm RR}$ and the CD is negligible.
The situation changes at M$_{\rm P2}$. RCP illumination redistributes the multipolar amplitudes and their relative phases with respect to LCP illumination, and the two vector sums in Figure~\ref{fig:plasmonic}d no longer terminate at the same point. This leads to the clear separation between $T_{\rm RR}$ and $T_{\rm LL}$ and to the pronounced CD observed at M$_{\rm P2}$.
The plasmonic-only metasurface therefore already contains the essential chiral seed required for the hybrid system.

\begin{figure*}[t]
    \centering
    \includegraphics[width=\linewidth]{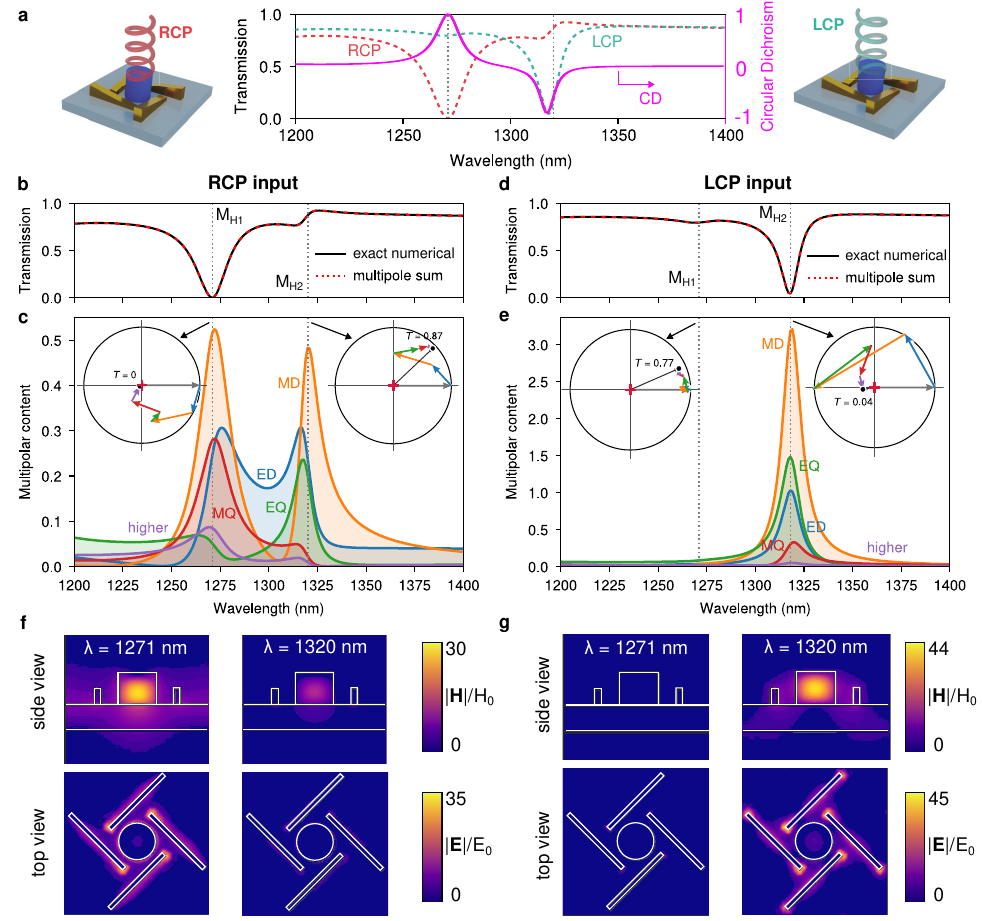}
    \caption{\textbf{Hybridized chiral response.}
    (a) Transmission of the plasmonic--dielectric hybrid metasurface under RCP and LCP illumination (left axis) and the resulting CD (right axis). At M$_{\rm H1}$ (1271~nm) the RCP channel is extinguished and ${\rm CD} \to +1$; at M$_{\rm H2}$ (1320~nm) the LCP channel is extinguished and ${\rm CD} \to -1$.
    (b,d) Transmission under RCP (b) and LCP (d) incidence, full-wave result (solid) and multipole sum \eqref{eq:t_multipole} (dotted). 
    (c,e) Multipolar content \eqref{eq:multipole_content}. Insets, complex-plane construction as in Fig.~\ref{fig:dielectric}c,d at M$_{\rm H1}$ and M$_{\rm H2}$. 
    (f,g) Near fields at the two resonances under RCP (f) and LCP (g), $|\mathbf H|/H_0$ in the vertical cut (top row) and $|\mathbf E|/E_0$ in the horizontal cut (bottom row). The magnetic field sits in the Si nanocylinder and the electric field at the Au nanobars, and both are strong only for the selected input helicity.}
    \label{fig:hybrid}
\end{figure*}

We now turn to the full plasmonic-dielectric hybrid metasurface in Figure~\ref{fig:hybrid} to determine how these two modal subsystems reorganize once they are combined into the same periodic meta-atom. 
Figures~\ref{fig:hybrid}c and~\ref{fig:hybrid}e show the multipolar content [Eq.~\eqref{eq:multipole_content}] under RCP and LCP 
illumination, respectively. Under LCP excitation the response is overwhelmingly concentrated at M$_{\rm H2}$, where the MD component dominates and is accompanied by pronounced ED and EQ contributions, with only weak excitation amplitudes around M$_{\rm H1}$. Under RCP excitation both resonances are driven with comparable amplitudes, all of them several times smaller than the LCP peak at M$_{\rm H2}$, as the different vertical scales of the two panels show. Both resonances preserve an overall MD-dominated character, but with appreciable electric and higher-order additions. The hybridization therefore does not simply replace one plasmonic mode with a dielectric mode. It reorganizes the same set of electric- and magnetic-type multipolar channels into two resonant states whose excitation strengths and phases depend strongly on the incident helicity.

How this content reaches the far field is shown by the complex-plane insets of Figures~\ref{fig:hybrid}c and~\ref{fig:hybrid}e. 
Under RCP illumination the multipolar vectors nearly cancel the direct term at M$_{\rm H1}$, giving $T \approx 0$, while at M$_{\rm H2}$ they leave $T = 0.87$. Under LCP illumination the roles are exchanged, with $T = 0.77$ at M$_{\rm H1}$ and $T = 0.04$ at M$_{\rm H2}$. 
Figures~\ref{fig:hybrid}f and~\ref{fig:hybrid}g show near-field maps.
The magnetic field is preferentially confined within the Si nanocylinder and the silica membrane, whereas the electric field is concentrated predominantly along the Au nanobars.

The handedness-dependent reorganization of these hybrid modes is translated directly into the observable transmission and CD spectra in Figure~\ref{fig:hybrid}a. At M$_{\rm H2}$, which lies close to the original plasmonic chiral resonance M$_{\rm P2}$, coupling to the Si Mie mode markedly amplifies the pre-existing helicity selectivity. The LCP transmission is nearly extinguished while the RCP channel stays finite, yielding a negative CD of $-0.91$. 
M$_{\rm H1}$ emerges from a spectral region where the plasmonic-only structure exhibits essentially no chiral response. Hybridization activates a strongly RCP-selective resonance, driving the RCP transmission to zero while leaving the LCP channel highly transmissive and thereby producing a positive CD peak approaching $+1$ (here the cross-polarized transmittance components are zero due to the preserved C$_4$ symmetry).
The resulting CD sign reversal is consequently an intrinsic consequence of the opposite helicity preferences of the two plasmonic-Mie hybrid resonances.

To verify the multipolar approach itself, we reconstruct the transmission from the VSH coefficients (Eq.\eqref{eq:expansion}) through Eq.~\eqref{eq:t_multipole} and compare it with the direct full-wave spectra (Figures~\ref{fig:hybrid}b and~\ref{fig:hybrid}d). 
The multipole sum reproduces both transmission minima, and hence the near-unity positive and negative CD extrema of Figure~\ref{fig:hybrid}a. The agreement confirms that the expansion is converged in $\ell$ and that the selection rules and normalization used in Eq.~\eqref{eq:t_multipole} are implemented correctly, so that the complex-plane construction accounts for the full transmitted field rather than a subset of it. The same reconstruction is shown for the dielectric-only structure in Figure~\ref{fig:dielectric}a and for the plasmonic-only structure in Figure S1 of the Supporting Information.

The Si-cylinder height and radius were varied independently while the plasmonic geometry was kept fixed (see Figure S2 of the Supporting Information) in order to identify the influence of the relative position and phase between interacting resonances. 
Both sweeps show a continuous evolution of the two chiral spectral branches, together with a pronounced redistribution and occasional sign reversal of the CD. These trends indicate that the chiral response is controlled by the evolution of the plasmonic-Mie coupling. Changing the dielectric dimensions alters the modal detuning, spatial overlap, and relative phase between the coupled constituents, which in turn modifies the helicity-dependent interference in transmission. The optimized geometry can therefore be viewed as a specific operating point at which this interference becomes nearly destructive for one circular polarization but not for the other, resulting in the observed near-unity CD.

\begin{figure*}[t]
	\centering
	\includegraphics[width=0.7\linewidth]{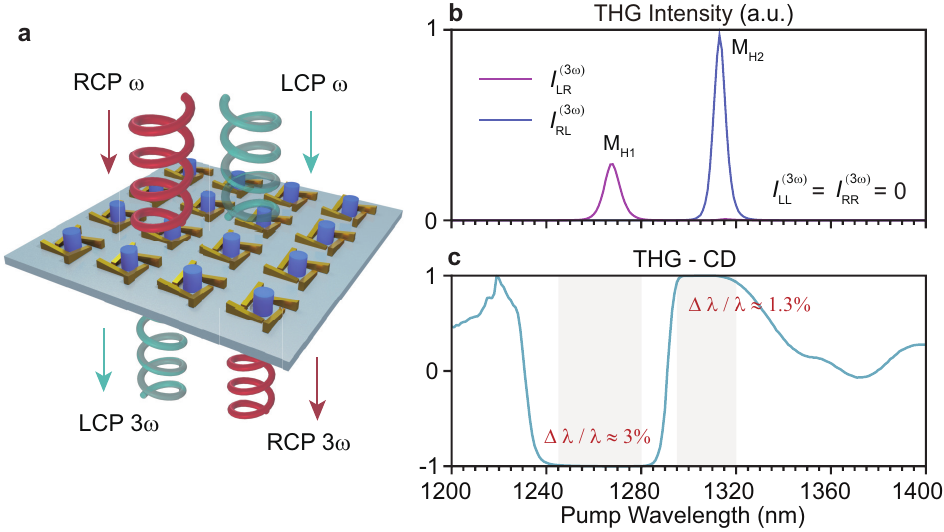}
	\caption{\textbf{Nonlinear chiral response.} (a) Schematic of the nonlinear chiral responses, illustrating the THG emission from the hybrid metasurface under RCP and LCP excitation. (b) Polarization-resolved THG spectra of the plasmonic--dielectric hybrid metasurface under circularly polarized pumping, showing the four THG intensities $I^{(3\omega)}_{\rm LL}$, $I^{(3\omega)}_{\rm RR}$, $I^{(3\omega)}_{\rm LR}$ and $I^{(3\omega)}_{\rm RL}$. (c) THG-CD spectrum calculated from the total THG powers under opposite circularly polarized pumping, showing near-unity extrema and sign reversal between the two resonances.}
	\label{fig:nonlinear}
\end{figure*}

Having established the polarization-selective plasmonic-Mie hybrid modes for linear chiroptical effects, we finally examine how it is transferred into nonlinear regime (Figure~\ref{fig:nonlinear}a).
Figure~\ref{fig:nonlinear}b reveals two sharply separated nonlinear resonances that follow the handedness selection of the underlying hybrid modes. Around  M$_{\rm H1}$, RCP pumping activates a strong THG response, dominated by the cross-circular THG emitted intensity $I^{(3\omega)}_{\rm LR}$, 
whereas the LCP-pumped channels are strongly suppressed. Conversely, near M$_{\rm H2}$, the nonlinear emission switches to the LCP-pumped pathway, with $I^{(3\omega)}_{\rm RL}$ becoming dominated, while rejecting the opposite handedness by orders of magnitude. Notably, the nonlinear symmetric selection rule \cite{Chen2014THGSelection,Koshelev2024NonlinearSMatrix} directs the THG emission mainly into the cross-circular channels, with the co-circular components being forbidden. The same normalization was performed for all channels when comparing LCP and RCP pumping.

This extreme pump-handedness selectivity is directly reflected in the third harmonic generation circular dichroism (THG-CD) spectrum in Figure~\ref{fig:nonlinear}c.
Under the ideal $C_4$ constraint,
$I^{(3\omega)}_{\rm LL}=I^{(3\omega)}_{\rm RR}=0$, the THG-CD reduces to~\cite{Petralli-Mallow1993J.Phys.Chem.,Toftul2024,Kim2020NanoLett.,Frizyuk2021NanoLett.,Tang2020LaserPhotonicsRev.}
\begin{equation}
    \mathrm{CD}^{(3\omega)}
    =
    \frac{
    I^{(3\omega)}_{\rm RL}-I^{(3\omega)}_{\rm LR}
    }{
    I^{(3\omega)}_{\rm RL}+I^{(3\omega)}_{\rm LR}
    }.
\end{equation}
Instead of producing isolated resonant extrema, $\mathrm{CD}^{(3\omega)}$ is pinned to $-1$ over a spectral window ($\sim$37 nm wide) near M$_{\rm H1}$ and close to $+1$ around M$_{\rm H2}$ ($\sim$17 nm wide), so that the two sign-reversed extrema are flat rather than pointlike. In fractional terms these windows are $\Delta\lambda/\lambda \approx 3\%$ and $1.3\%$. This differs from resonance-enhanced nonlinear chirality, where a large nonlinear CD appears only at isolated spectral points. 
Here, the saturation originates from the near-complete extinction of one circular pumping pathway over an extended wavelength range, while the opposite pathway continues to dominate the THG output. Therefore, the nonlinear response provides a higher-contrast readout of the Mie-assisted enhancement of chiral response. The same plasmonic--Mie hybridization that produces sign-reversing, high-purity linear CD is converted into a near-unity, sign-switchable THG-CD response.

In conclusion, we have demonstrated a plasmonic-dielectric hybrid metasurface in which dielectric Mie resonators actively mediate, amplify, and reverse the intrinsic chiral response of plasmonic structures. 
Isolation analyses reveal that the dielectric component provides achiral but spectrally defined Mie scattering channels, while the plasmonic component supplies a finite chiral seed. When integrated into the full hybrid structure, these two ingredients are reorganized into polarization-selective hybrid modes. Unlike a simple superposition of plasmonic and dielectric resonances, the hybrid meta-atom supports two closely spaced electric/magnetic hybrid modes formed through plasmonic--Mie coupling. As a result, the Mie-mediated hybridization does not merely enhance the plasmonic CD, but reorganises the original chiral plasmonic response into two near-unity CD extrema of opposite sign at adjacent hybrid modes.
Being extended to the nonlinear regime, the same hybrid mechanism underpins near-unity, sign-reversible THG-CD over finite spectral windows rather than at isolated points. These results suggest the Mie-assisted plasmonic hybridization as a powerful route for engineering high-purity and sign-switchable chiroptical responses in both linear and nonlinear regimes, providing a compact platform for chiral sensing, polarization-selective and nonlinear frequency conversion metadevices.

\medskip
\noindent\textit{Supporting Information.}
The Supporting Information contains details of the numerical simulation, vector-spherical-harmonic multipole,
transmission reconstruction methods and COMSOL implementation of the VSHs calculation.

\begin{acknowledgments}
The authors thank the support of the National Natural Science Foundation of China (NSFC) (Grant No.12304417); Natural Science Foundation of Heilongjiang Province of China (Grant No.YQ2024A007) and the Fundamental Research Funds for the Central Universities (Grant No.XNJKKGYDJ2026019). 
\end{acknowledgments}

\bibliography{refs}

\end{document}